# A multivariate stochastic model to assess research performance[1]

*Giovanni Abramo[a], Corrado Costa[b], Ciriaco Andrea D'Angelo[c]*

[a] Laboratory for Studies of Research and Technology Transfer
at the Institute for System Analysis and Computer Science (IASI-CNR)
National Research Council of Italy

[b] Consiglio per la Ricerca e la sperimentazione in Agricoltura (CRA)
Unità di ricerca per l'Ingegneria Agraria

[c] Department of Engineering and Management
University of Rome "Tor Vergata"

**Abstract**

There is a worldwide trend towards application of bibliometric research evaluation, in support of the needs of policy makers and research administrators. However the assumptions and limitations of bibliometric measurements suggest a probabilistic rather than the traditional deterministic approach to the assessment of research performance. The aim of this work is to propose a multivariate stochastic model for measuring the performance of individual scientists and to compare the results of its application with those arising from a deterministic approach. The dataset of the analysis covers the scientific production indexed in Web of Science for the 2006-2010 period, of over 900 Italian academic scientists working in two distinct fields of the life sciences.

## 1. Introduction

Bibliometric instruments are increasingly applied in support of research assessment of individuals and institutions, due to their advantages as objective measures in comparing the performance of different entities. Research assessment serves many purposes, among others: informing research policies at the national and supranational levels; informing strategies at the organizational level; stimulating research productivity; supporting selective funding; reducing asymmetry in demand and supply of new knowledge; demonstrating that investment in research is effective and delivers public benefits. Because of the delicate purposes served by research assessment exercises, policy makers and research administrators require ever more accurate, robust, timely, and inexpensive measures.

Not surprisingly, recent years have seen proliferation of different bibliometric indicators, often with further variations and ever more sophisticated methods for their calculation. The assumptions and the limits of application for bibliometrics, and of the individual indicators and their combinations in research evaluation, are such as to suggest the consideration of a probabilistic rather than the current deterministic approach to the measurement of research performance. The aim of this work is to propose a multivariate model for the measurement of individual research performance, and to compare the results with those derived from a deterministic approach. We expect that randomness will be particularly significant for evaluation exercises conducted at the level of individual scientists, but decreasing as the levels of analysis are increasingly aggregated (research groups, departments, universities).

The next section of the study illustrates the principal factors that determine uncertainty in the measurement of research performance, while Section 3 presents the performance indicators adopted, the multivariate model, the field of observation and the measurement methodology. Sections 4 and 5 present the results of the application and a comparison to the results that would derive from a traditional deterministic approach. The paper closes with the authors' considerations and some suggestions for future developments in the research.

## 2. Uncertainty factors in research performance assessment

One of the major areas of difficulty in measuring research performance concerns the multi-output character of the research produced. Bibliometrics is not able to measure any new knowledge that is not codified, and where new knowledge is indeed codified, bibliometricians are still faced with the problem of identifying and measuring its various forms. Moed (2005) demonstrated that in the so-called hard sciences, the prevalent form of codification for research output is publication in scientific journals. Given this, databases such as Scopus and Web of Science (WoS) have been extensively used and tested in bibliometric analyses, and are seen as sufficiently transparent in terms of their content and coverage. As a proxy of total output in the hard sciences, bibliometricians thus simply consider publications indexed in either WoS or Scopus[2].

---

[2] Although the overall coverage achieved by the two databases does differ significantly, evidence suggests that with respect to comparisons at large scale level in the hard sciences, the use of either source yields similar results (Archambault et al. 2009).

The immediate consequence is that those outputs that are not censused will inevitably be ignored. This approximation is considered acceptable in the hard sciences, although not for the arts, humanities and a good part of the social sciences.

Publications embedding new knowledge have different values, measured by their impact on scientific advancement. As proxy of this impact, bibliometricians adopt the number of citations of the publications (in spite of limits on the indicator, such as negative citations and "network" citations) (Glänzel 2008). Because citation behavior varies across fields, bibliometricians then standardize the citations of each publication with respect to a scaling factor stemming from the distribution of citations for all publications of the same year and the same subject category. Different scaling factors have been suggested and adopted for the field normalization of citations (average, median, z-score of normalized distributions, etc.). Because interdisciplinary work may easily suffer in the evaluation from being misplaced in a categorical classification system (Laudel & Origgi 2006), few scholars have proposed to normalize citations by the number of bibliographic references of the citing paper (Pepe & Kurtz 2012; Leydesdorff & Bornmann 2011). However it can be expected that no parametric field-normalization method will be able to realize the perfect overlapping of citation distributions in the case of a substantial number of fields (Zhang et al. 2014; Abramo et al. 2012a; Radicchi et al. 2008). Furthermore, given the varying intensity of publications across fields (Butler 2007; Garfield 1979), in order to avoid distortions in performance rankings (Abramo et al. 2008), evaluation exercises should compare researchers within the same field. A prerequisite of any research assessment free of distortions would then be the classification of each individual researcher in one and only one field. An immediate corollary is that the performance of staff units that are heterogeneous for fields of research cannot be directly measured at the aggregate level, and evaluators must apply a two-step procedure: first measuring the performance of the individual researchers in their field, and then appropriately aggregating this data. The classification of researchers in fields, while absolutely necessary, is not an easy task, and thus in itself embeds uncertainty.

Other approximations and limits are seen to apply for the individual bibliometric indicators. Abramo & D'Angelo (2014), for example, compare the strengths and weaknesses of the most popular performance indicators, such as the new crown indicator, the h-index, and fractional scientific strength. The choice of indicators and measurement methods cannot be divorced from the objectives and context of the assessment exercise, however most bibliometricians agree on the need to adopt some form of combination of indicators for the evaluation of individuals and institutions, rather than a single instrument. Among the pros of composite indicators, are the facts that they: i) are easier to interpret than a battery of many separate indicators; ii) make it possible to include more information; iii) enable users to compare complex dimensions effectively. Among the points against are: i) the weighting processes applied in the combination of variables are arbitrary in nature and may lead to disputes; ii) they may send misleading policy messages if poorly constructed or misinterpreted; iii) they may lead to simplistic management or policy conclusions. Combining different indicators is not an exercise to take lightly, given the number of hidden dangers involved. Without at least minimal knowledge of aggregation techniques and properties, as applied in multi-criteria decision-making (MCDM), evaluators risk the vitiation of their entire exercise. A case in point is the Academic Ranking of World Universities, popularly known as the "Shanghai ranking", released annually by the Institute of Higher Education, Jiao Tong

University of Shanghai. In adding an MCDM analysis to van Raan's (2005) previous bibliometric critique, Billaut et al. (2010) quickly came to the conclusion that the Shanghai ranking is a poorly conceived, "quick and dirty" exercise. They observe that the method used to aggregate indicators is flawed and nonsensical, and that no attention has been paid to fundamental structuring issues.

Bibliometric scores are also afflicted by further factors of randomness. Referring to the "publication window", bibliometricians have observed that both the time period from a paper's original submission to a journal and its date of acceptance, and then from acceptance to actual publication date, are highly variable within the same discipline. This means that the shorter the observation period assessed, the greater the citation measures will be affected by a random component external to the excellence of the researchers. Publication delays have been noted as particularly evident in certain fields, such as mathematics and technical sciences (Luwel & Moed 1998), food sciences (Amat 2008) and econometrics (Trivedi 1993). Further, the intensity of a scientist's publication production is clearly linked to the type of research taken on, to whether it is more or less innovative or in "niche" areas, and to the entire research life cycle: a scientist could result as completely unproductive if evaluated during the launch of a new research program, while his or her performance would be completely different if evaluated in the subsequent stages of "harvesting" the yields from initial investments. Another frequently noted temporal effect relates to the length of the "citation window", another source of randomness in bibliometric scores. The reliability of citations to approximate the real impact of a publication is clearly higher the wider is the time window for citation.

The decision of any author to cite or not to cite an article is in itself a stochastic process. Once cited, the indexing of citations on the part of the major database operators then relies on algorithms, and although such indexing processes are constantly improved they will never be free of error. The indexing itself typically takes place at the beginning or end of a calendar year, so that in measuring impact there is bias in favor of publications published earlier.

We must also admit that the very probability of acceptance for a paper submitted to a journal is affected by random factors related to interactions of individuals in the reviewing process of the manuscript, and by the fact that the choice of the journal by the authors, and of the reviewers by the editors, is not always optimal.

In spite of the numerous elements of randomness illustrated, traditional bibliometric assessments are largely based on deterministic models which perform the same way for a given set of initial conditions. Conversely, in a stochastic model, randomness is present, and variable states are not described by unique values, but rather by probability distributions. The literature does provide several studies of the statistical properties of bibliometric indicators, relative to the research performance of individuals (Radicchi & Castellano 2013), research groups (van Raan 2006) and universities as a whole (van Raan 2008). Concerning the widest known of all indicators, the h-index, the large part of the contributions are dedicated to identifying its weaknesses, while only a few studies focus on its statistical properties (Cerchiello & Giudici 2014; Pratelli et al. 2012; Todeschini 2011; Burrel 2007).

## 3. Methods and data

The current study proposes a composite indicator for the evaluation of research performance at the individual level. For this, in order to filter the effects of randomness on the bibliometric scores, we adopt a stochastic rather than deterministic approach. The approach used is Soft Independent Modeling of Class Analogy (SIMCA) (Vanden Branden & Hubert 2005; Wold & Sjostrom 1977). For a rigorous description of the method in detail we refer the interested reader to the above works. Here we attempt to provide a synopsis of the method which may be comprehensible to the average reader.

SIMCA is a multivariate statistical method for supervised classification of data. To avoid over-fitting, the method requires a training data set consisting of samples with a set of attributes and their class membership. The term soft refers to the fact the classifier can identify samples as belonging to multiple classes and not necessarily producing a classification of samples into non-overlapping classes. In order to build the classification models, the samples belonging to each class need to be analyzed using principal components analysis (PCA). For a given class, the resulting model then describes either a line (for one Principal Component or PC), plane (for two PCs) or hyper-plane (for more than two PCs). For each modeled class, the mean orthogonal distance of training data samples from the line, plane or hyper-plane (calculated as the residual standard deviation) is used to determine a critical distance for classification. This critical distance is based on the F-distribution and is usually calculated using 95% or 99% confidence intervals. New observations are projected into each PC model and the residual distances calculated. An observation is assigned to the model class when its residual distance from the model is below the statistical limit for the class. The observation may be found to belong to multiple classes and a measure of goodness of the model can be found from the number of cases where the observations are classified into multiple classes. The classification efficiency is usually indicated by receiver operating characteristics.

In the original SIMCA method, the ends of the hyper-plane of each class are closed off by setting statistical control limits along the retained principal components axes (i.e. range: minimum score value minus 0.5 times score standard deviation to maximum score value plus 0.5 times standard deviation). More recent adaptations of the SIMCA method close off the hyper-plane by construction of ellipsoids. With such modified SIMCA methods (Forina et al. 2008a), classification of an object requires both that its orthogonal distance from the model and its projection within the model (*i.e.*, score value within region defined by ellipsoid) are not significant.

SIMCA as a method of classification has gained widespread use especially in applied statistical fields such as chemometrics and spectroscopic data analysis (Menesatti et al. 2014; Aguzzi et al. 2009; Forina et al. 2008a, 2008b; Casale et al. 2007; Hall & Kenny 2007; Krafft et al. 2006). To the best of our knowledge this is the first application of SIMCA to the fields of bibliometrics and research evaluation. In fact, this type of multivariate approach is particularly suited for treatment of correlated data that could present stochastic fluctuations (Forina et al. 2008a), as in the case of our bibliometric indicators.

Our SIMCA application, computed with the software V-Parvus 2010, is based on a multivariate dataset composed of five bibliometric indicators that describe the performance of scientists from different points of view. The first indicator, Fractional Scientific Strength (FSS) (Abramo & D'Angelo 2014), is an indicator of efficiency that measures the research productivity of the subjects evaluated, accounting for both quantity and impact of production. Two of the indicators represent excellence, in terms of the number of publications of the author that place in the top 1% and 5% of world ranking for impact, referring to works of the same subject category and year. The last two indicators refer to the relative importance of the contribution by the authors of the co-authored works, which is particularly important in the life sciences fields that are the object of our particular application. These indicators are based on counts of the number of publications in which the scientist appears as first or last author in the byline. The SIMCA model is calibrated on two artificial datasets constructed from the basis of the performance distributions of all Italian professors belonging to the fields of pharmacology (506 observations) and general pathology (417 observations). The bibliometric dataset is specifically based on their scientific production as indexed in the Web of Science, over the period 2006 to 2010.

In the following subsections we provide more detailed descriptions of the performance indicators adopted, the multivariate model, the field of observation and the evaluation methodology.

**3.1 The performance indicators**

Any performance should be evaluated relative to goals and objectives as stated for the given context. Because objectives will necessarily vary across research institutions and over time, the recommendation of a sole performance indicator is inappropriate. However this does not on the other hand justify the proliferation of hundreds of indicators. In this work we propose five bibliometric indicators. Two of these are specifically for those fields where the varying contributions of the co-authors are signaled through their order in the article's byline.

Bibliometric measurement of research performance requires the adoption of a number of simplifications and assumptions. In the current work, as in others, one of the most delicate of these is that the same resources are available to all scientists in the same field.

In the vast majority of evaluation exercises, the first and likely most important indicator of performance is research productivity. The current study measures this by the indicator named Fractional Scientific Strength (FSS), which embeds both the number of publications produced and their standardized impact.

Because the intensity of publications varies across fields, in order to avoid distortions in productivity rankings, we compare researchers within the same field. A prerequisite of any productivity assessment free of distortions is then a classification of each individual researcher in one and only one field. We take advantage of a unique feature of the Italian higher education system, in which each professor is classified as belonging to a single research field. These formally-defined fields are called Scientific Disciplinary Sectors (SDSs): there are 370 SDSs, grouped into 14 University Disciplinary Areas (UDAs). In the hard sciences, there are 205 such fields[3] grouped into nine UDAs[4]. We then compare the performance of all professors belonging to the same

---

[3] The complete list is accessible on http://attiministeriali.miur.it/UserFiles/115.htm, last accessed on September 15, 2014.
[4] Mathematics and computer sciences; physics; chemistry; earth sciences; biology; medicine; agricultural and veterinary sciences; civil engineering; industrial and information engineering.

SDS. In formula, the average yearly productivity of an individual, over a period of time, is[5]:

$$FSS_R = \frac{1}{t}\sum_{i=1}^{N}\frac{c_i}{\bar{c}}f_i$$

[1]

Where:
t = number of years of work of the professor in the period of observation;
N = number of publications of the professor in the period of observation;
$c_i$ = citations received by publication *i*;
$\bar{c}$ = average of the distribution of citations received by all cited publications[6] of the same year and subject category of publication *i*;
$f_i$ = fractional contribution of the researcher to publication *i*.

We adopt the methodology of fractional counting of research contributions, as we believe it is more compatible with microeconomic theory of production than "full counting". The methodology permits accounting to the level of the different contributions of individual authors, where this is signaled by their position in the byline. Fractional contribution equals the inverse of the number of authors, in those fields where the practice is to place the authors in simple alphabetical order, but assumes different weights in other cases. For the life sciences, widespread practice in Italy and abroad is for the authors to indicate the various contributions to the published research by the order of the names in the byline. For these disciplines, we give different weights to each co-author according to their order in the byline and the character of the co-authorship (intra-mural or extra-mural). If first and last authors belong to the same university, 40% of citations are attributed to each of them; the remaining 20% are divided among all other authors. If the first two and last two authors belong to different universities, 30% of citations are attributed to first and last authors; 15% of citations are attributed to second and last author but one; the remaining 10% are divided among all others[7]. Failure to account for the number and position of authors in the byline would result in notable ranking distortions at the individual level (Abramo et al. 2013a).

While productivity is the quintessential indicator of efficiency in any production system, another important indicator of performance is research excellence, i.e. the ability to achieve path-opening discoveries. We thus measure, for each professor, the number of articles that rank among the top 1% ($HCA_{1\%}$) and 5% ($HCA_{5\%}$) world publications (of the same year and subject category[8]) per number of citations. Finally, in the life sciences, the byline entry of the first author of the publication generally indicates the generator of the original concept, as well as the scientist who contributed the most to the research and writing. Correspondingly, the position of last author generally indicates the team manager of the research project. Being either first or last author is a sign of distinction and is highly acknowledged in the scientific world. We then measure the number of articles where a professor is either the first (First$_A$) or the last author (Last$_A$). In general, due to the particularities of the Italian context, we can exclude that performance measures may be distorted by variable returns to scale, due to different sizes of universities (Abramo et al. 2012b) or by returns to the differing scope of their research fields (Abramo et al. 2013b).

---

[5] A thorough description of the formula, the underlying theory, assumptions and limits is found in Abramo & D'Angelo (2014).
[6] A preceding article by the same authors demonstrated that the average of the distribution of citations received for all cited publications of the same year and subject category is the most effective scaling factor (Abramo et al. 2012a).
[7] The weighting values were assigned following advice from prestigious Italian representatives of the scientific community in the life sciences. The values could be changed to suit different practices in other national contexts.
[8] When articles are published in multidisciplinary journals we assign them to the subject category where they rank the highest.

## 3.2 Data

The field of observation consists of all Italian professors belonging to two fields (SDSs) of the life sciences: Pharmacology (BIO/14, 506 observations) and General pathology (MED/04, 417 observations). Data on professors and their SDS classification are extracted from the database on Italian university personnel, maintained by the Ministry for Universities and Research[9]. For the bibliometric dataset, we draw on the Italian Observatory of Public Research (ORP), a database developed and maintained by the authors and derived under license from the WoS. Beginning from the raw data of Italian publications indexed in WoS, and applying a complex algorithm for disambiguation of the true identity of the authors and their institutional affiliations (for details see D'Angelo et al. 2011), each publication is attributed to the university professor that produced it, with a harmonic average of precision and recall (F-measure) equal to 96 (error of 4%). The observation period is 2006-2010. Once the five-year scientific portfolio is reconstructed for the 923 professors in the dataset, for each of these we measure the five indicators described in section 2.1. Table 1 presents the descriptive statistics for the distributions of the indicators, relative to the two SDSs under examination.

*Table 1: Descriptive statistics of performance indicators measured for Italian professors of the dataset*

| SDS | Index | Mean | Median | Min | Max | Std Dev. | Variat. coeff. | Skewness | Kurtosis |
|---|---|---|---|---|---|---|---|---|---|
| BIO/14 (506 obs.) | FSS | 2,75 | 1,44 | 0 | 46,23 | 4,23 | 1,54 | 4,59 | 31,47 |
| | First$_A$ | 1,53 | 1 | 0 | 23 | 2,21 | 1,45 | 3,58 | 23,91 |
| | Last$_A$ | 3,81 | 2 | 0 | 94 | 6,26 | 1,64 | 6,80 | 85,15 |
| | HCA$_{1\%}$ | 0,17 | 0 | 0 | 8 | 0,65 | 3,82 | 6,14 | 51,90 |
| | HCA$_{5\%}$ | 0,85 | 0 | 0 | 13 | 1,65 | 1,95 | 3,15 | 12,95 |
| MED/04 (417 obs.) | FSS | 3,20 | 1,12 | 0,01 | 98,79 | 6,90 | 2,16 | 7,74 | 90,56 |
| | First$_A$ | 1,30 | 0 | 0 | 25 | 2,38 | 1,83 | 4,29 | 29,12 |
| | Last$_A$ | 3,87 | 2 | 0 | 48 | 5,74 | 1,48 | 3,56 | 18,17 |
| | HCA$_{1\%}$ | 0,30 | 0 | 0 | 10 | 0,98 | 3,31 | 5,18 | 34,39 |
| | HCA$_{5\%}$ | 1,21 | 0 | 0 | 39 | 3,00 | 2,48 | 6,61 | 65,83 |

## 3.3 Multivariate modeling

Multivariate class-modeling techniques answer to the general question of whether an object O, stated of class A, really belongs to class A (Forina et al. 2008a; Taiti et al., 2014). This is a typical question in multivariate quality control. On the contrary, traditional classification techniques assign objects to one, and only one, of the potential classes. Class-modeling techniques calculate the "prediction probability" with a classification threshold for each modeled class. In using a class-modeling approach it is possible to attribute objects to one or more classes, but also to none, meaning the object is observed as an outlier.

---
[9] http://cercauniversita.cineca.it/php5/docenti/cerca.php, last accessed on September 15, 2014

To develop an index suitable for assessing the research performance of individual researchers, we apply the SIMCA model to the five bibliometric performance indicators, as described above. For the modeled class, a critical squared distance based on the F-distribution is calculated using 95% confidence interval (*i.e.*, the class boundary). The efficiency is indicated by a classification (training set) and a prediction (evaluation set) matrix, which report percentage of correct classification for each considered class and the total percentage ability. The observations for each class classified outside the model are also reported. SIMCA expresses the statistical parameters indicating the modeling efficiency. In fact, the observation can be found to belong to multiple classes or to fit none of them (outlier). Also, unknown objects could be either classified into one of the classes or recognized as outliers. The modeling efficiency is indicated by a sensitivity parameter. The sensitivity is the measure of how well the classification test correctly identifies the observations really belonging to the class, thus providing a quantitative indication of how well the model was capable of correctly classifying the researchers. The modeling power of each variable is also expressed, representing the influence of that particular variable in the definition of the model.

To express an index for each researcher, squared SIMCA distances are linearized by converting the values into logarithmic scale and then translating them, adding a value of 1 (to result in all positive values) to both datasets (BIO/14 and MED/04). This index expresses not only if researchers fall or not in the model (artificial dataset; see the following section) depending on whether its value is below or above the modeled threshold respectively, but also the quantitative proximity to the threshold (*i.e.*, the relative performance of the researcher).

*Training*

The modeling approach used is based on the construction of artificial observations, characterized by absolute excellence from the point of view of bibliometric performance. This construction is based on the cut of the upper tail of the performance distribution for the five bibliometric indicators measured on the true dataset, as per Table 1. Specifically, for each SDS analyzed we construct an artificial dataset by means of a full permutation design. Given the distribution of values for the five performance indicators measured, we identify the four "high" performance values (the $95^{th}$ percentile of the distribution; the $97.5^{th}$ percentile; the maximum value; the maximum value increased by 5%) and "construct" 1,024 artificial observations by means of their factorial combination (4 values for 5 indicators, *i.e.* $4^5$).

To avoid over-fitting, of the 1,024 artificial observations only 75% (the training and validation sets) are used to construct and cross-validate the SIMCA model

*Testing*

The remaining 25% of the artificial dataset is now used to test the performance of the SIMCA model. The partitioning of the artificial datasets is optimally chosen with Euclidean distances, based on the Kennard & Stone (1969) algorithm that selects objects without *a priori* knowledge of a regression model (*i.e.*, the hypothesis is that a flat distribution of the data is preferable for a regression model).

Once the SIMCA models are constructed and tested on the artificial datasets, they are then run through with the real datasets, based on the true performance of the professors in the two SDSs.

To proceed to the comparative analysis between the results derived from the application of the stochastic SIMCA model and those of a deterministic approach, we also calculate a "traditional" composite indicator called Bibliometric Composite Score (BCS), given by the weighted average of the standardized values for the five indicators. The weights applied are FSS, 50%; $HCA_{1\%}$, 20%; $HCA_{5\%}$, 10%; $First_A$, 10%; $Last_A$, 10%[10]. The individual values of the indicators are standardized to the mean value of the distribution of all Italian professors of the same SDS with values above 0.

## 4. The research performance by the multivariate model

Figure 1 presents the performance of the SIMCA model, calibrated for BIO/14 (Pharmacology). The figure shows the histogram for frequency classes of the translated log squared SIMCA. The critical value results as 1.58 (red dashed line). Below this value, the observations are classified as belonging to the reference model, while observations with a value above threshold are considered "external to the model". The greater is the value of the translated log squared SIMCA distance, the greater is the distance from the reference model, and thus from bibliometric excellence. We observe that all the observations used both for training (blue bars) and for testing the model (red bars) are "acceped" (100% sensitivity), while of the 506 true professors of the BIO/14, 62 (12.3%) show values above the threshold. Of these 62, 47 (76% of total) belong to the best 50 as identified using the "traditional" BCS. The PCA shows the presence of two principal components, with a re-computed class standard deviation (RSD) equal to 1.21.

The Figure 2 histogram shows the modeling power, or the weight that each of the five variables has in the SIMCA model score. We observe that the greatest contributions are from FSS and from $HCA_{1\%}$.

---

[10] The weights applied may be changed according to the specific objectives of the evaluation.

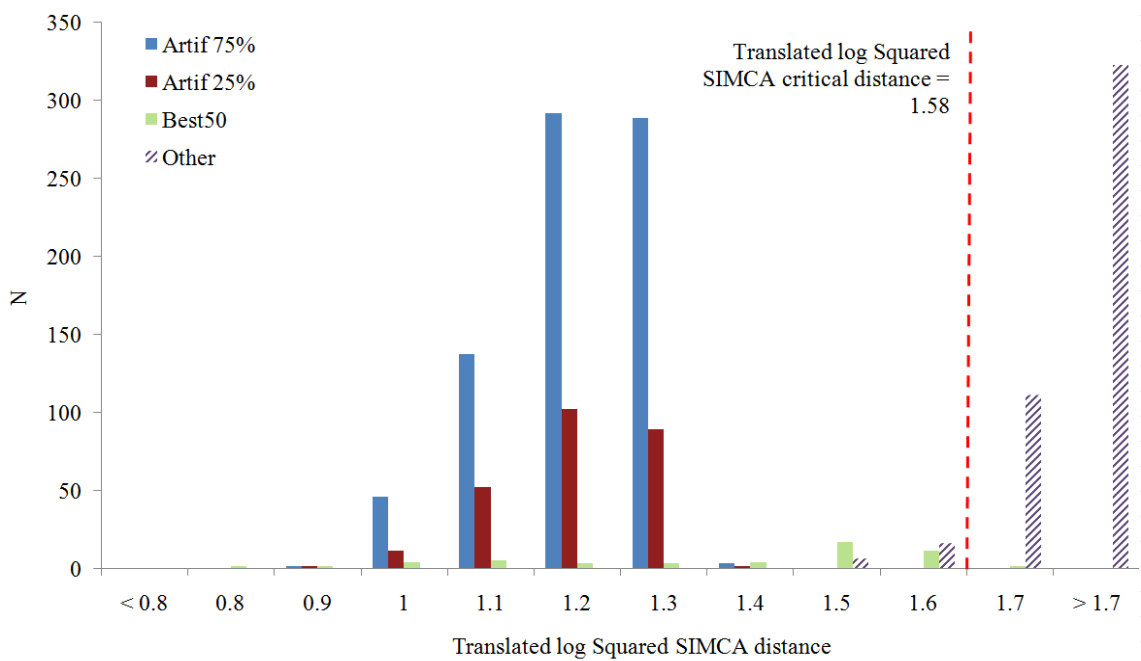

*Figure 1: Histogram by frequency class of the translated log squared SIMCA for observations of BIO/14 (Pharmacology), subdivided in four groups: i) the 75% of the artificial dataset used to build the model (Artif75%; blue); ii) the 25% artificial datasets used as external test (Artif25%; red); iii) the best 50 real researchers identified using the "deterministic" BCS (Best50; green); iv) all other real researchers (Other; purple). The dashed red line is the critical value.*

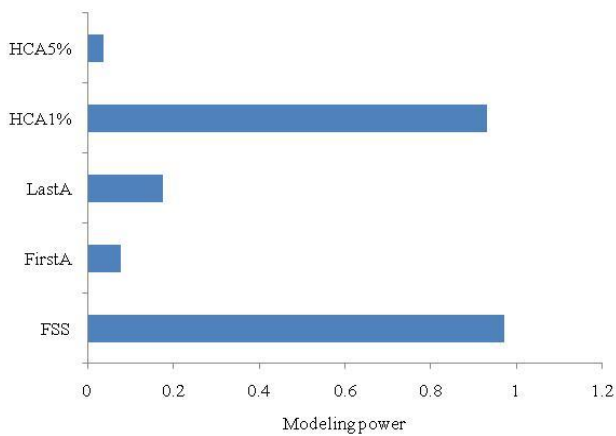

*Figure 2: Modeling power of the 5 bibliometric indicators in the SIMCA model for BIO/14*

Figure 3 presents the performance of the SIMCA model calibrated for MED/04 (General pathology). The translated log squared SIMCA critical distance is equal to 1.60. The entirety of artificial observations used to calibrate the model (blue bars) and test it (red bars) show values above this threshold, indicating a sensitivity level of 100%. On the other hand, of the 417 real professors, 99 (23.7%) are accepted by the model. These include all of the top 50 professors as identified by BCS. The remaining 318 (purple bars) are rejected, falling outside the "excellent" class. The PCA shows the presence of two principal components and a re-computed class standard deviation (RSD) equal to 1.23.

The Figure 4 histogram shows the modeling power of the five bibliometric indicators in the SIMCA model. The graph is virtually identical to Figure 3: once again, the greatest weight in the model is from FSS and $HCA_{1\%}$.

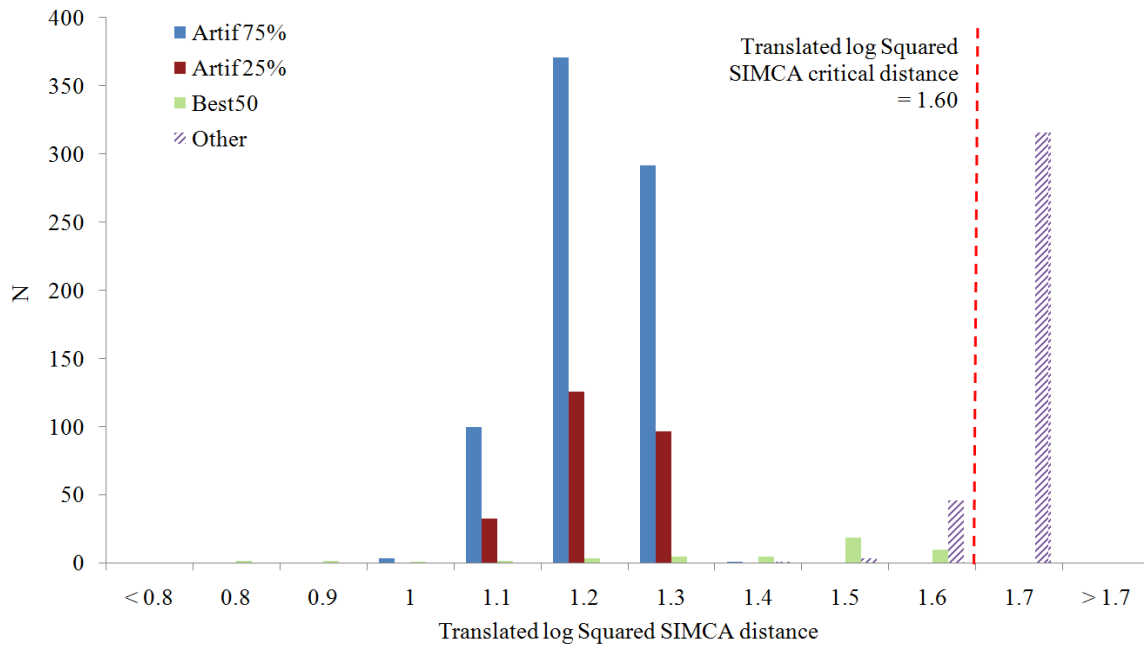

*Figure 3: Histogram by frequency class of the translated log squared SIMCA for observations of MED/04 (General pathology), subdivided in four groups: i) the 75% of the artificial dataset used to build the model (Artif75%; blue); ii) the 25% artificial datasets used as external test (Artif25%; red); iii) the best 50 real researchers identified using BCS (Best50; green); iv) all other real professors (Other; purple). The critical value is indicated as a dashed red line.*

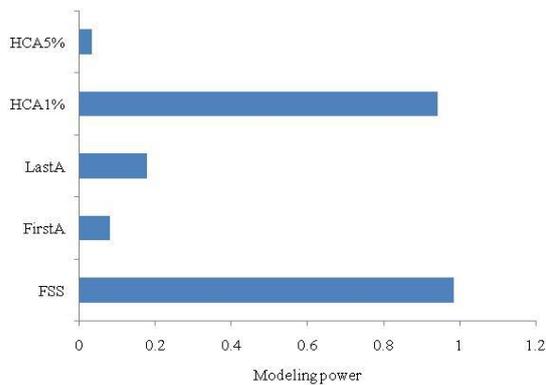

*Figure 4: Modeling power of the 5 bibliometric indicators to the SIMCA model for MED/04*

## 5. Comparison between the multivariate model and the deterministic approach

Table 2 shows the bibliometric performance of the 50 professors of BIO/14 that are top ranked by BCS. The last two columns show the relative scores by SIMCA, permitting comparison between the BCS score and the SIMCA distance. Forty-two of the top-BCS professors also result as top scientists on the basis of SIMCA classification. Of the eight excluded professors, the first is ID 125, who places 26$^{th}$ in the ranking for BCS. For this individual, the SIMCA distance is particularly remarkable, placing the person completely outside the best 50 professors of the SDS. We observe in particular that the professor in question never appears as first or last author of articles published. The same lack of first/last authorship occurs for ID 179, however this professor has authored a full 18 top-1% articles.

The most apparent outlier is professor ID 139, who shows the top-placing BCS, more than 50% greater than the second-ranked professor (ID 300). This second professor in fact registers a log squared SIMCA that is slightly less than for his higher-ranking colleague (1.035 vs. 1.071).

For the same BIO/14 professors, Figure 5 shows the scatter plot of regressions between the BCS and SIMCA rankings. The Spearman correlation ($\rho = 0.95$) is clearly very high, although the graph also shows some dispersion.

*Table 2: Bibliometric performance of the top 50 BIO/14 professors by BCS, with relative SIMCA score*

| Researcher ID | FSS | $HCA_{1\%}$ | $HCA_{5\%}$ | $First_A$ | $Last_A$ | BCS | SIMCA score |
|---|---|---|---|---|---|---|---|
| R_139 | 46.228 | 23 | 94 | 1 | 8 | 11.412 | 1.071 Best 20 |
| R_300 | 31.506 | 5 | 12 | 8 | 13 | 7.438 | 1.035 Best 10 |
| R_163 | 27.347 | 0 | 16 | 3 | 11 | 6.033 | 0.928 Best 10 |
| R_19 | 26.703 | 1 | 5 | 4 | 5 | 5.589 | 1.020 Best 10 |
| R_316 | 20.915 | 0 | 13 | 4 | 10 | 4.868 | 0.911 Best 10 |
| R_279 | 21.929 | 2 | 16 | 3 | 4 | 4.833 | 1.013 Best 10 |
| R_154 | 19.019 | 2 | 10 | 2 | 7 | 4.235 | 0.958 Best 10 |
| R_102 | 17.009 | 4 | 17 | 2 | 8 | 4.114 | 0.831 Best 10 |
| R_391 | 15.911 | 0 | 6 | 4 | 7 | 3.702 | 0.984 Best 10 |
| R_162 | 15.333 | 1 | 26 | 1 | 5 | 3.642 | 1.181 Best 20 |
| R_30 | 16.461 | 1 | 14 | 2 | 3 | 3.633 | 1.184 Best 20 |
| R_369 | 13.003 | 6 | 12 | 4 | 8 | 3.555 | 0.795 Best 10 |
| R_334 | 14.826 | 4 | 21 | 0 | 5 | 3.490 | 1.279 Best 20 |
| R_317 | 15.547 | 0 | 21 | 1 | 2 | 3.427 | 1.342 Best 20 |
| R_200 | 14.607 | 1 | 30 | 0 | 2 | 3.366 | 1.428 Best 30 |
| R_50 | 12.406 | 0 | 5 | 3 | 6 | 2.925 | 1.111 Best 20 |
| R_476 | 12.142 | 0 | 19 | 0 | 6 | 2.861 | 1.356 Best 20 |
| R_311 | 11.053 | 6 | 0 | 2 | 7 | 2.774 | 1.060 Best 10 |
| R_179 | 10.481 | 18 | 2 | 0 | 0 | 2.683 | 1.520 Best 50 |
| R_14 | 10.624 | 5 | 5 | 1 | 6 | 2.614 | 1.217 Best 20 |
| R_143 | 10.645 | 0 | 25 | 0 | 2 | 2.523 | 1.497 Best 50 |
| R_231 | 9.96 | 1 | 13 | 1 | 6 | 2.482 | 1.287 Best 20 |
| R_439 | 10.66 | 2 | 12 | 0 | 2 | 2.372 | 1.487 Best 40 |
| R_77 | 10.598 | 0 | 13 | 1 | 1 | 2.348 | 1.467 Best 40 |
| R_174 | 10.002 | 0 | 11 | 0 | 5 | 2.283 | 1.432 Best 30 |
| R_125 | 9.71 | 2 | 20 | 0 | 0 | 2.256 | 1.568 |
| R_489 | 8.35 | 4 | 15 | 0 | 5 | 2.211 | 1.415 Best 30 |
| R_83 | 8.679 | 2 | 15 | 0 | 4 | 2.152 | 1.457 Best 40 |
| R_466 | 8.534 | 0 | 25 | 0 | 2 | 2.140 | 1.535 |
| R_142 | 9.178 | 2 | 6 | 1 | 2 | 2.079 | 1.439 Best 30 |
| R_16 | 8.699 | 4 | 2 | 1 | 3 | 2.041 | 1.395 Best 30 |
| R_198 | 9.2 | 0 | 9 | 1 | 1 | 2.024 | 1.497 Best 50 |
| R_441 | 8.647 | 7 | 1 | 0 | 2 | 2.001 | 1.496 Best 50 |
| R_312 | 8.068 | 1 | 3 | 2 | 4 | 1.969 | 1.346 Best 20 |
| R_325 | 7.785 | 0 | 11 | 0 | 6 | 1.932 | 1.459 Best 40 |
| R_460 | 7.996 | 3 | 11 | 0 | 2 | 1.907 | 1.528 |
| R_378 | 7.241 | 0 | 15 | 2 | 2 | 1.903 | 1.440 Best 30 |
| R_168 | 7.488 | 0 | 15 | 0 | 4 | 1.859 | 1.506 Best 50 |
| R_209 | 7.256 | 0 | 12 | 1 | 4 | 1.854 | 1.435 Best 30 |
| R_199 | 8.747 | 0 | 0 | 1 | 2 | 1.824 | 1.481 Best 40 |
| R_450 | 6.847 | 0 | 16 | 0 | 4 | 1.761 | 1.520 Best 50 |
| R_236 | 7.463 | 4 | 10 | 0 | 0 | 1.746 | 1.600 |
| R_112 | 8.139 | 1 | 3 | 0 | 2 | 1.714 | 1.552 |
| R_194 | 6.868 | 3 | 6 | 0 | 4 | 1.701 | 1.499 Best 50 |
| R_99 | 6.161 | 0 | 12 | 2 | 3 | 1.699 | 1.438 Best 30 |
| R_420 | 6.684 | 3 | 11 | 0 | 2 | 1.674 | 1.555 |
| R_225 | 6.628 | 2 | 18 | 0 | 0 | 1.665 | 1.623 |
| R_478 | 6.45 | 6 | 1 | 0 | 4 | 1.653 | 1.491 Best 40 |
| R_1 | 7.105 | 2 | 7 | 0 | 1 | 1.597 | 1.590 |
| R_203 | 6.427 | 2 | 6 | 1 | 2 | 1.586 | 1.504 Best 50 |

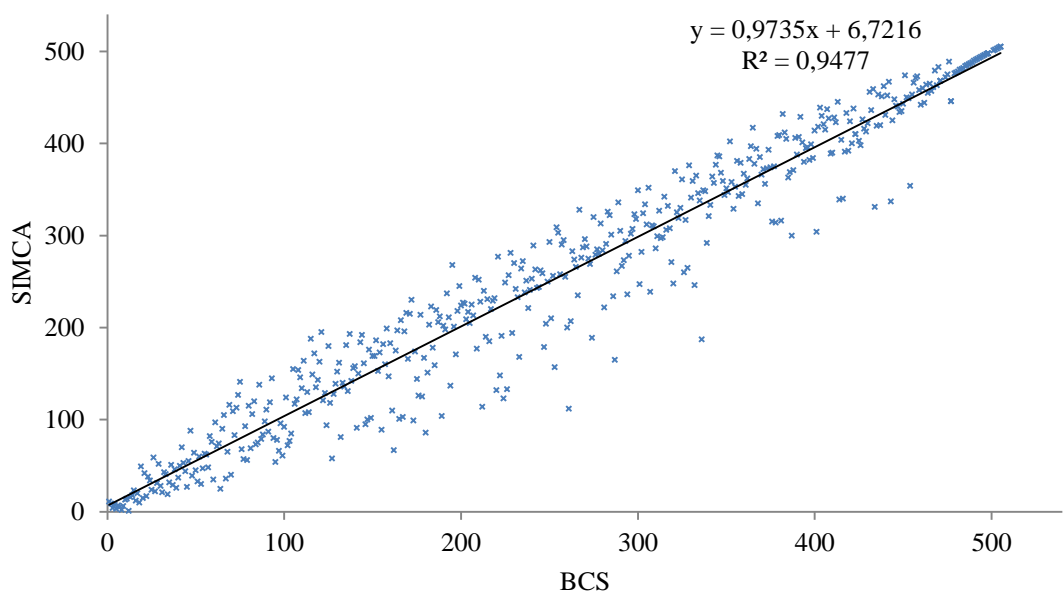

**Figure 5: Scatter plot of BCS vs SIMCA rankings for BIO/14 professors**

Table 3 shows the bibliometric performance of the top 50 professors of MED/04, by BCS. The last two columns permit comparison of the BCS scores and SIMCA distances. Forty-three of the top BCS professors also result as top scientists on the basis of SIMCA classification. Of the seven professors excluded, ID 174 places 26[th] in the BCS ranking. For this individual, the SIMCA distance is particularly relevant, placing the professor completely outside the best 50 of the SDS. We note in particular that the researcher in question did not author any top 1% articles over the study period. However while they did not publish any top 1% articles, professors ID 118, 409 and 55 still remain in the top 50 by SIMCA score, having scored higher than ID 174 for the other indicators. In this case too, the top-ranked professor for BCS, ID 128 is an evident outlier. For him/her the BCS score is more than double that of the second-ranked professor.

Figure 6 shows the plot of regressions between the BCS and SIMCA rankings for the MED/04 professors. Here again there is a very high and statistically significant Spearman correlation ($\rho = 0.94$), although the graph also shows a meaningful dispersion.

*Table 3: Bibliometric performance of top 50 MED/04 professors by BCS, with relative SIMCA score*

| Researcher ID | FSS | HCA$_{1\%}$ | HCA$_{5\%}$ | First$_A$ | Last$_A$ | BCS | SIMCA score |
|---|---|---|---|---|---|---|---|
| R_128 | 98.794 | 10 | 39 | 9 | 35 | 18.461 | 1.025 | Best 10 |
| R_213 | 34.858 | 3 | 18 | 5 | 24 | 6.935 | 0.757 | Best 10 |
| R_180 | 33.658 | 7 | 17 | 1 | 14 | 6.674 | 0.898 | Best 10 |
| R_71 | 32.948 | 6 | 14 | 5 | 12 | 6.504 | 0.773 | Best 10 |
| R_380 | 27.304 | 6 | 12 | 2 | 37 | 5.904 | 0.837 | Best 10 |
| R_165 | 27.644 | 4 | 15 | 2 | 12 | 5.452 | 0.926 | Best 10 |
| R_92 | 26.825 | 2 | 13 | 3 | 7 | 5.059 | 1.157 | Best 20 |
| R_189 | 23.199 | 3 | 5 | 1 | 48 | 4.983 | 1.131 | Best 10 |
| R_132 | 26.482 | 4 | 6 | 0 | 7 | 4.808 | 1.130 | Best 10 |
| R_123 | 20.067 | 4 | 10 | 0 | 10 | 3.988 | 1.122 | Best 10 |
| R_153 | 18.24 | 2 | 5 | 1 | 37 | 3.928 | 1.232 | Best 20 |
| R_130 | 21.361 | 3 | 7 | 0 | 4 | 3.915 | 1.231 | Best 20 |
| R_126 | 20.246 | 1 | 7 | 0 | 12 | 3.745 | 1.364 | Best 20 |
| R_188 | 14.808 | 3 | 5 | 14 | 13 | 3.511 | 1.066 | Best 10 |
| R_120 | 18.741 | 1 | 7 | 1 | 8 | 3.474 | 1.376 | Best 20 |
| R_178 | 19.383 | 1 | 6 | 0 | 6 | 3.468 | 1.402 | Best 30 |
| R_118 | 15.392 | 0 | 4 | 1 | 39 | 3.344 | 1.459 | Best 40 |
| R_11 | 10.919 | 2 | 5 | 25 | 8 | 3.142 | 1.291 | Best 20 |
| R_409 | 13.469 | 0 | 3 | 6 | 24 | 2.924 | 1.432 | Best 30 |
| R_194 | 12.122 | 5 | 10 | 0 | 12 | 2.865 | 1.201 | Best 20 |
| R_360 | 14.058 | 1 | 7 | 0 | 12 | 2.775 | 1.410 | Best 30 |
| R_22 | 14.017 | 1 | 3 | 1 | 13 | 2.690 | 1.416 | Best 30 |
| R_161 | 12.566 | 1 | 8 | 1 | 11 | 2.592 | 1.408 | Best 30 |
| R_369 | 11.398 | 1 | 1 | 5 | 21 | 2.506 | 1.383 | Best 30 |
| R_55 | 11.188 | 0 | 3 | 2 | 19 | 2.327 | 1.499 | Best 50 |
| R_174 | 11.71 | 0 | 1 | 1 | 17 | 2.270 | 1.520 | |
| R_359 | 10.522 | 0 | 4 | 0 | 23 | 2.254 | 1.514 | Best 50 |
| R_238 | 10.842 | 0 | 4 | 1 | 16 | 2.215 | 1.514 | Best 50 |
| R_195 | 8.901 | 3 | 7 | 8 | 1 | 2.211 | 1.276 | Best 20 |
| R_141 | 9.266 | 2 | 3 | 1 | 14 | 2.043 | 1.381 | Best 20 |
| R_316 | 9.002 | 1 | 5 | 2 | 7 | 1.907 | 1.459 | Best 40 |
| R_125 | 10.136 | 0 | 2 | 2 | 4 | 1.853 | 1.555 | |
| R_387 | 7.603 | 3 | 5 | 5 | 2 | 1.851 | 1.333 | Best 20 |
| R_147 | 8.326 | 2 | 3 | 0 | 11 | 1.797 | 1.416 | Best 30 |
| R_340 | 9.544 | 1 | 4 | 0 | 2 | 1.793 | 1.503 | Best 50 |
| R_315 | 8.719 | 0 | 5 | 0 | 10 | 1.771 | 1.553 | |
| R_334 | 7.946 | 1 | 4 | 2 | 10 | 1.764 | 1.463 | Best 40 |
| R_58 | 8.109 | 0 | 3 | 3 | 10 | 1.722 | 1.537 | |
| R_191 | 8.591 | 0 | 3 | 1 | 9 | 1.703 | 1.555 | |
| R_179 | 8.148 | 1 | 4 | 3 | 2 | 1.686 | 1.483 | Best 40 |
| R_15 | 6.048 | 0 | 0 | 15 | 3 | 1.616 | 1.534 | |
| R_18 | 5.646 | 1 | 1 | 11 | 9 | 1.615 | 1.439 | Best 30 |
| R_311 | 7.058 | 1 | 4 | 5 | 3 | 1.608 | 1.471 | Best 40 |
| R_373 | 7.302 | 1 | 4 | 0 | 9 | 1.571 | 1.496 | Best 50 |
| R_91 | 6.911 | 1 | 5 | 0 | 8 | 1.521 | 1.498 | Best 50 |
| R_185 | 6.158 | 2 | 6 | 0 | 7 | 1.491 | 1.437 | Best 30 |
| R_346 | 6.424 | 1 | 4 | 1 | 7 | 1.429 | 1.500 | Best 50 |
| R_262 | 6.633 | 0 | 4 | 0 | 9 | 1.393 | 1.576 | |
| R_394 | 6.099 | 1 | 3 | 5 | 1 | 1.387 | 1.494 | Best 50 |
| R_309 | 5.522 | 2 | 7 | 0 | 2 | 1.333 | 1.461 | Best 40 |

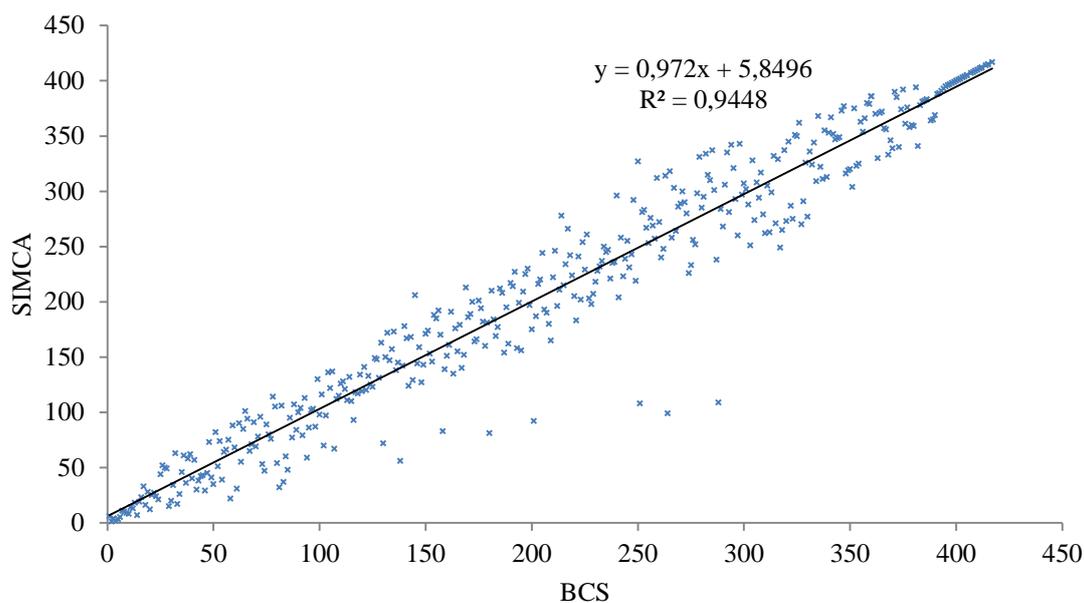

**Figure 6:** Scatter plot of BCS vs SIMCA rankings for MED/04 professors

## 6. Discussion

Bibliometric measures of research performance are based on assumptions, which determine important limits that must be kept in account when measurements are used to support policy and research administration decisions. Besides failing to capture production of new knowledge that is not codified in publication or where the publication is not indexed, bibliometric indicators are affected by many other random factors. Some of these are related to the actual processes of publishing research, others to the initial assumptions involved in measuring and comparing the performance of individuals and research units.

To hedge against the randomness of these undetected effects, as well as to evaluate different dimensions of research performance, bibliometricians often resort to a combination of indicators in the evaluation of research, rather than just one. However the combination of individual indicators again introduces other elements of randomness to the performance evaluation.

This leads to the suggestion of a stochastic approach to research evaluation, whatever the bibliometric indicator or set of indicators that may be used. For this, the current paper has proposed a multivariate class-modeling approach, which has been developed in different fields and applied to various research problems, particularly in biomedicine and chemometrics. In these fields the variables under study, and their relative values, are typically correlated between each other and strongly affected by chance, hence the statistical approach results as more effective than deterministic ones (Costa et al. 2012). It was precisely the analogy to these contexts that suggested the possible application of stochastic techniques to bibliometric evaluation. In this particular paper we have used the SIMCA multivariate indicator for the estimation of the individual research performance of over 900 Italian university scientists working in two life sciences fields, over the 2006-2010 period. The approach is based on five indicators that measure different aspects of research performance. The SIMCA score then measure the "distance" of each evaluated individual from a "class" characterized by absolute excellence in the five indicators. This class is constructed in an artificial manner, considering the possible combinations from the upper tail of the distributions of the five indicators.

The SIMCA rank results as being highly correlated with a "deterministic" composite indicator composed of the weighted values of the same five starting indicators, thus indicating a convergence in the results from the two approaches. Notwithstanding the high correlation between ranks, few individuals register significant jumps about 15% of the top-50 scientists by the SIMCA method are not as such by the deterministic approach. Since there is no unequivocal reference benchmark, we cannot know which of the two approaches in fact provides a more truthful representation of the real value of the scientists. However we can reason that, given the stochastic nature of the variables in play, the SIMCA multivariate class-modeling approach would be a valid substitute to the deterministic approach precisely in those real contexts where bibliometric indicators will be affected by higher randomness, such as in cases where the publication window or citation window are restricted.

Although randomness should be particularly significant for evaluation exercises conducted at the level of individual scientists, but decrease with the increasing aggregation of larger units of analysis, the authors believe that a worthwhile future aspect of this research would be apply the stochastic method to the evaluation of research teams, departments and entire institutions. Furthermore, in addition to bibliometric indicators, where available other indicators of scientific merit such as patents, licenses, spin-off companies, attraction of research funds and the likes could be considered in the SIMCA model.

**References**

Abramo, G., D'Angelo, C.A., & Di Costa, F. (2008). Assessment of sectoral aggregation distortion in research productivity measurements. *Research Evaluation*, 17(2), 111-121.

Abramo, G., Cicero, T., & D'Angelo, C.A. (2012a). Revisiting the scaling of citations for research assessment. *Journal of Informetrics*, 6(4), 470–479.

Abramo, G., Cicero, T., & D'Angelo, C.A. (2012b). Revisiting size effects in higher education research productivity. *Higher Education*, 63(6), 701-717.

Abramo, G., D'Angelo, & C.A., Rosati, F. (2013a). The importance of accounting for the number of co-authors and their order when assessing research performance at the individual level in the life sciences. *Journal of Informetrics,* 7(1), 198–208.

Abramo, G., D'Angelo, C.A., & Di Costa, F. (2013b). Investigating returns to scope of research fields in universities. *Higher Education.* DOI: 10.1007/s10734-013-9685-x

Abramo, G., & D'Angelo, C.A., (2014). How do you define and measure research productivity? *Scientometrics.* DOI: 10.1007/s11192-014-1269-8

Aguzzi, J., Costa, C., Antonucci, F., Company, J. B., Menesatti, P., & Sardá, F. (2009). Influence of diel behaviour in the morphology of decapod natantia. *Biological Journal of the Linnean Society*, 96, 517-532.

Amat, C.B. (2008). Editorial and publication delay of papers submitted to 14 selected Food Research journals. Influence of online posting. *Scientometrics*, 74(3), 379-

389.

Archambault, É., Campbell, D., Gingras, Y., & Larivière, V. (2009). Comparing bibliometric statistics obtained from the Web of Science and Scopus. *Journal of the American Society for Information Science and Technology*, 60(7), 1320-1326

Billaut, J.C, Bouyssou, D., & Vincke, P. (2010). Should you believe in the Shanghai ranking? An MCDM view. *Scientometrics,* 84, 237–263.

Burrell, Q.L. (2007). Hirsch's h-index: A stochastic model. *Journal of Informetrics*, 1, 16–25.

Butler, L. (2007). Assessing university research: A plea for a balanced approach. *Science and Public Policy*, 34(8), 565-574.

Casale, M., Armanino, C., Casolino, C., & Forina, M. (2007). Combining information from headspace mass spectrometry and visible spectroscopy in the classification of the Ligurian olive oils. *Analytica chimica acta*, 589(1), 89–95.

Cerchiello, P., & Giudici, P. (2014). On a statistical h index. *Scientometrics*, 99, 299-312.

Costa, C., Menesatti, P., & Spinelli, R. (2012). Performance modelling in forest operations through partial least square regression. *Silva Fennica*, 46(2), 241-252.

D'Angelo, C.A., Giuffrida C., & Abramo, G. (2011). A heuristic approach to author name disambiguation in bibliometrics databases for large-scale research assessments. *Journal of the American Society for Information Science and Technology*, 62(2), 257-269.

Forina, M., Oliveri, P., Casale, M., & Lanteri, S. (2008b). Multivariate range modeling, a new technique for multivariate class modeling: The uncertainty of the estimates of sensitivity and specificity. *Analytica chimica acta*, 622(1), 85-93.

Forina, M., Oliveri, P., Lanteri, S., & Casale, M. (2008a). Class-modeling techniques, classic and new, for old and new problems. *Chemometrics and Intelligent Laboratory Systems*, 93(2), 132–48.

Garfield, E. (1979). Is citation analysis a legitimate evaluation tool? *Scientometrics,*1(4), 359-375.

Glänzel, W. (2008). Seven myths in bibliometrics. About facts and fiction in quantitative science studies. Kretschmer, & F. Havemann (Eds): *Proceedings of WIS Fourth International Conference on Webometrics, Informetrics and Scientometrics, & Ninth COLLNET Meeting*, Berlin, Germany.

Hall, G.J., & Kenny, J.E. (2007). Estuarine water classification using EEM spectroscopy and PARAFAC-SIMCA. *Analytica chimica acta*, 581(1), 118–24.

Kennard, R.W., & Stone, L.A. (1969). Computer aided design of experiments. *Technometrics*, 11(1), 137–48.

Krafft, C., Shapoval, L., Sobottka, S.B., Geiger K.D., Schackert G., & Salzer R. (2006). Identification of primary tumors of brain metastases by SIMCA classification of IR spectroscopic images. *Biochimica et Biophysica Acta (BBA)-Biomembranes*, 1758(7), 883-891.

Laudel, G., & Origgi, G. (2006). Introduction to a special issue on the assessment of interdisciplinary research. *Research Evaluation*, 15(1), 2–4.

Leydesdorff, L., & Bornmann, L. (2011). How fractional counting of citations affects the impact factor: Normalization in terms of differences in citation potentials among fields of science. *Journal of the American Society for Information Science and Technology*, 62(2), 217-229.

Luwel, M., & Moed, H.F. (1998). Publication delays in the science field and their relationship to the ageing of scientific literature. *Scientometrics,* 41(1-2), 29-40.

Menesatti, P., Antonucci, F., Pallottino, F., Bucarelli, F.M., & Costa, C. (2014). Spectrophotometric qualification of Italian pasta produced by traditional or industrial production parameters. *Food and Bioprocess Technology*, 7(5), 1364-1370

Moed, H.F. (2005). *Citation Analysis in Research Evaluation*. Springer, ISBN: 978-1-4020-3713-9.

Pepe, A., & Kurtz, M.J. (2012). A Measure of total research impact independent of time and discipline. *PLoS ONE,* 7(11), e46428.

Pratelli, L., Baccini, A., Barabesi, L., & Marcheselli, M., (2012). Statistical analysis of the Hirsch index. *Scandinavian Journal of Statistics*, 39, 681–694.

Radicchi, F., & Castellano, C. (2013). Analysis of bibliometric indicators for individual scholars in a large data set. *Scientometrics*, 97 (3), 627-637.

Radicchi, F., Fortunato, S., & Castellano, C. (2008). Universality of citation distributions: Toward an objective measure of scientific impact. *Proceedings of the National Academy of Sciences of the United States of America,* 105(45), 17268-17272.

Taiti, C., Costa, C., Menesatti, P., Comparini, D., Bazihizina, N., Azzarello, E., Masi, E., & Mancuso, S. (2014). Class-modeling approach to PTR-TOFMS data: a peppers case study. Journal of the Science of Food and Agriculture (accepted on 22/05/2014)

Todeschini, R. (2011). The j-index: A new bibliometric index and multivariate comparisons between other common indices. *Scientometrics*, 87, 621–639.

Vanden Branden, K., & Hubert, M. (2005). Robust classification in high dimensions based on the SIMCA method. *Chemometrics and Intelligent Laboratory Systems,* 79(1-2), 10-21.

van Raan, A.F.J., (2005), Fatal attraction: Conceptual and methodological problems in the ranking of universities by bibliometric methods. *Scientometrics*, 62(1),133–143.

van Raan, A.F.J. (2006). Statistical properties of bibliometric indicators: Research group indicator distributions and correlations. *Journal of the American Society for Information Science and Technology*, 57 (3), 408-430.Trivedi, P.K. (1993). An analysis of publication lags in econometrics. *Journal of Applied Econometrics*, 8(1), 93-100.

van Raan, A.F.J. (2008). Bibliometric statistical properties of the 100 largest European research universities: Prevalent Scaling rules in the science system. *Journal of the American Society for Information Science and Technology*, 59 (3), 461-475.

Wold, S., & Sjostrom, M., (1977). SIMCA: A method for analyzing chemical data in terms of similarity and analogy. In *Chemometrics: Theory and Application*, Kowalski, B.R., ed., American Chemical Society Symposium Series 52, Wash., D.C., p. 243-282.

Zhang, Z., Cheng, Y., & Liu, N.C., (2014). Comparison of the effect of mean-based method and z-score for field normalization of citations at the level of Web of Science subject categories. *Scientometrics*, DOI 10.1007/s11192-014-1294-7